\documentclass[twocolumn]{mn2e}
\newif\ifAMStwofonts
\usepackage{ulem}
\usepackage{color}

\def\pmb#1{\mbox{\boldmath$#1$}}
\def\gtsim {>\kern-1.2em\lower1.1ex\hbox{$\sim$}}
\def\ltsim {<\kern-1.2em\lower1.1ex\hbox{$\sim$}}
\def\gtsim {>\kern-1.2em\lower1.1ex\hbox{$\sim$}}
\def\ltsim {<\kern-1.2em\lower1.1ex\hbox{$\sim$}}
\def\ref{\hangindent=1pc \hangafter=1 \noindent}
\def\be{\begin{equation}}
\def\ee{\end{equation}}

\usepackage{epsfig}
\begin{document}

\title{Angular momentum transport by stochastically excited oscillations
in rapidly rotating massive stars}
\author[U. Lee, C. Neiner, S. Mathis]{Umin Lee$^1$\thanks{E-mail: lee@astr.tohoku.ac.jp},
Coralie Neiner$^2$\thanks{E-mail: Coralie.Neiner@obspm.fr}, and St\'ephane Mathis$^3$\thanks{E-mail: stephane.mathis@cea.fr}
\\$^1$Astronomical Institute, Tohoku University, Sendai, Miyagi 980-8578, Japan
\\$^2$LESIA, UMR 8109 du CNRS, Observatoire de Paris, UPMC, Univ. Paris Diderot,
5 place Jules Janssen, 92195 Meudon Cedex, France
\\$^3$Laboratoire AIM Paris-Saclay, CEA/DSM-Universit\'e Paris Diderot-CNRS, IRFU/SAp Centre de Saclay, 91191
Gif-sur-Yvette, France}

\date{Typeset \today ; Received / Accepted}
\maketitle


\begin{abstract}
We estimate the amount of angular momentum transferred by the low-frequency
oscillations detected in the rapidly rotating {hot} Be star
HD\,51452. Here, we assume that the oscillations detected are stochastically
excited by convective motions in the convective core of the star, that is,  we
treat the oscillations as forced oscillations excited by the  periodic
convective motions of the core fluids having the frequencies observationally
determined. With the observational amplitudes of the photometric variations, we
determine the  oscillation amplitudes, which makes it possible to estimate the
net amount of angular momentum transferred by the oscillations using the 
wave-meanflow interaction theory. Since we do not have any information
concerning the azimuthal wavenumber $m$ and spherical harmonic degree $l$ for
each of the oscillations, %
{we assume} that all the frequencies
detected are prograde or retrograde in the observer's frame and they are all
associated with a single value of $m$ both for even modes ($l=|m|$) and for odd
modes ($l=|m|+1$). We estimate the amount of angular momentum transferred by the
oscillations for $|m|=1$ and 2{, which are typical $|m|$ values for
Be stars,}  and find that the amount is large enough for a decretion disc to
form around the star.  %
{Therefore, transport of angular
momentum by waves stochastically excited in the core of Be stars might be
responsible for the Be  phenomenon.}
\end{abstract}


\section{Introduction}

Be stars are rapidly rotating {active} late O, B, or early A stars hosting a
circumstellar decretion disc {fed} by discrete mass loss
events. {The} mass ejections and {disc} produce  emission lines
in the optical spectrum of Be stars (see, e.g., Porter \& Rivinius 2003;
Rivinius et al. 2013 for recent reviews on Be stars), and {it
is believed} that the circumstellar discs around Be stars are viscous Keplerian
discs  (Lee, Saio {\&} Osaki 1991). Be stars of late O and
early B type are also known as $p$-mode pulsators, while those of  late B and
early A type are $g$-mode pulsators, where both the $p$-mode and $g$-mode
pulsations  are excited by the opacity mechanism associated with the opacity
bump produced by iron-peak elements at the temperature regions of $T\sim
2\times10^5$K in their envelope (e.g., Pamyatnykh 1999). { In addition, the
convective core of intermediate-mass and massive stars are able to
{stochastically excite} oscillation modes (e.g., Belkacem,
Dupret, Noels 2010; Samadi et al.. 2010, Shiode et al. 2013).
{Propagative gravity (gravito-inertial in the case of rapid
rotation) waves and modes are excited (e.g., Browning, Brun, Toomre 2004; Rogers
et al. 2013; Mathis, Neiner \& Tran Minh 2014), because of 
the convective motions and of their penetration into
the surrounding radiative envelope.} These waves are able to transport angular
momentum because of their dissipation along their propagation and at corotation
layers ({see} e.g., Zahn, Talon, Matias 1997 and Alvan, Mathis,
Decressin 2013 for gravity {waves,} and Pantillon et al. 2007
and Mathis et al. 2008 for gravito-inertial waves).}

Mechanisms for the mass ejections and disc formation, however, have not yet been
identified for Be stars. For disc formation mechanisms,
{several models have been proposed} such as stellar wind models
(e.g., Bjorkman \& Cassinelli 1993; Owocki, Cranmer, \& Blondin 1994; Cranmer \&
Owocki 1995), a model making use of magnetic fields to support a disc
(Cassinelli et al. 2002; Owocki \& ud Doula 2002), an evolutionary model that
assumes angular momentum redistribution in the interior of rotating stars (e.g.,
Ekstr\"om et al. 2008; Granada et al. 2013), and a model using
{$\kappa$-driven} pulsations as a carrier of angular momentum
to the surface region of rapidly rotating stars (e.g., Ando 1983, 1986; Lee \&
Saio 1993; Cranmer 2005, 2009; Ishimatsu \& Shibahashi 2013).  The stellar wind
models are not necessarily successful in producing Keplerian discs around the
stars. Magnetic fields strong enough to support circumstellar discs are not
common among Be stars { (e.g., Neiner et al. 2012b)}. On the other hand, {
as discussed above}, stellar pulsations in Be stars are excited by the opacity
bump mechanism or{/and} probably by stochastic excitation in the convective
core. In this sense using the pulsations as the angular momentum carrier can be
a promising mechanism for disc formation around Be stars, particularly when both
stellar evolution and pulsation play their own parts. In fact, as suggested by
Lee (2013), steady viscous decretion disc solutions are possible if a sufficient
amount of angular momentum is deposited {in} the surface
layers.

Recently, Neiner et al. ({2012a}) observed
{the} hot Be star HD\,51452 {with} the CoRoT
{satellite} and identified numerous low-frequency oscillations.
Since the star is rapidly rotating at the rate 1.22 cycle per day and the
oscillation frequencies in the observer's frame are comparable to or
{lower} than the rotation frequency, the oscillations are low-frequency ones in the corotating frame of the star, suggesting that the effects
of rapid rotation of the star can be significant to determine the wave
properties propagating in the star. It is also important to note that low-frequency 
oscillations  in early B type main sequence stars such as HD\,51452
cannot be excited by the opacity bump mechanism. Neiner et al.
({2012a}) therefore suggested that the low-frequency
oscillations detected are {gravito-inertial waves} excited stochastically by
the convective motions in the core of the star.

In this paper, we estimate a possible amount of angular momentum transferred by
the low-frequency oscillations identified by Neiner et al.
({2012a}) for the Be star HD\,51452. We calculate non-adiabatic
low-frequency oscillations by taking the effects of rotation of the star
{into account} assuming that the rotation is uniform (Lee \& Baraffe
1996), and we employ a wave-meanflow interaction theory (e.g., Andrews \&
McIntyre 1978ab; Grimshaw 1984; Lee 2013) to estimate the amount of angular
momentum transfer. The oscillations are treated as forced oscillations whose
frequencies are equal to those identified {with CoRoT} for the
star. Since the CoRoT {observations do} not provide any
information concerning the azimuthal wavenumber $m$ and spherical harmonic
degree $l$ for the oscillations, we employ in this paper a working hypothesis
that all the oscillations identified have the same $m$ and $l$ when estimating
the possible amount of angular momentum. We describe the
{analytical} method of calculation in \S2 and numerical results
in \S 3. Conclusions are given in \S 4.

{
\section{Theory for the transport of angular momentum}
}

{
\subsection{Wave-meanflow equation}
}

We discuss the angular momentum transfer by low-frequency waves excited by the
convective motions in the core of a massive star. To estimate the amount of
angular momentum transferred by the waves, we employ a theory of wave-meanflow
interaction, in which a wave-meanflow equation may be given in the Cowling
approximation by
\be
\rho {d\ell\over dt}=\sum_m{1\over 2}{\rm Im}\left[\nabla\cdot\left(m\pmb{\xi}_m^*p_m^\prime\right)\right],
\label{eq:wmeq}
\ee
where $\ell$ is the specific angular momentum in the $z$-direction along the
rotation axis, $m$ is the azimuthal wave number around the axis, $\pmb{\xi}$ is
the displacement vector, $p^\prime$ is the pressure Eulerian perturbation
associated with the waves, and the asterisk $(^*)$ indicates complex
conjugation. Here, {as a first step, our goal is to} estimate the amount of
angular momentum possibly carried by low-frequency oscillations to the surface
region when the oscillations are excited by periodic convective motions of the
fluids in the core. {The {completely} coupled time
evolution of both the meanflow and waves as a result of their interactions will
be studied in a forthcoming work.} We may represent the waves associated with
the azimuthal wavenumber $m$ as
\be
\pmb{\xi}_m=\sum_{\alpha}
\pmb{\xi}_{m\alpha}e^{i\omega_{m\alpha} t+i\delta_{m\alpha}}, 
\ee
\be
p^\prime_m=\sum_\alpha 
p^{\prime}_{{m\alpha}}e^{i\omega_{m\alpha} t+i\delta_{m\alpha}},
\ee
where $\omega_{m\alpha}=\sigma_{m\alpha}+m\Omega$, with $\Omega$ being the rotation frequency,
is the oscillation frequency observed in the corotating frame of the star, $\sigma_{m\alpha}$ is the frequency in an inertial (observer's) frame,
$\delta_{m\alpha}$
is the phase shift, where $\alpha$ indicates
the mode indices, for example, $\alpha=(l,n)$ with the degree $l$ and radial order $n$ for a mode
in a non-rotating star.
Note also that since we assume the time and azimuthal angle dependence of the perturbations are 
given by $\exp\left(i\omega t+i m\phi\right)$ or by $\exp\left(i\sigma t+i m\phi\right)$,
prograde and retrograde modes are respectively
associated with negative and positive values of $m$.
For a given value of $m$, the term $\pmb{\xi}_{m}^*p_m^\prime$ contains cross terms
$\pmb{\xi}_{m\alpha}^*p^\prime_{m\beta}e^{i\Delta\omega_{\beta\alpha}t+i\Delta\delta_{\beta\alpha}}$ for $\alpha\not=\beta$, where $\Delta\omega_{\beta\alpha}=\omega_{m\beta}-\omega_{m\alpha}$ and $\Delta\delta_{\beta\alpha}=\delta_{m\beta}-\delta_{m\alpha}$.
Since $e^{i\Delta\omega_{\beta\alpha}t}$ is rapidly oscillating except when
$\Delta\omega_{\beta\alpha}\simeq 0$, we ignore the terms with $\alpha\not=\beta$ in this paper.
In this approximation, we obtain
\be
\rho {d\ell\over dt}=\sum_{m,\alpha}{m\over 2}
{\rm Im}\left[\nabla\cdot\left(\pmb{\xi}_{m\alpha}^*p_{m\alpha}^\prime\right)\right].
\label{eq:alpbeta}
\ee

We take the effects of rotational deformation of the equilibrium structure on
the oscillations {into account}  by introducing the mean radius
$a$ of the equipotential surface defined as
\be
r=a\left[1+\epsilon(a,\theta)\right]
\ee
where $(r,\theta,\phi)$ are spherical polar coordinates, and $\epsilon(a,\theta)=
\alpha(a)+\beta(a)P_2(\cos\theta)$ with $P_2=(3\cos^2\theta-1)/2$ being a
Legendre polynomial is a quantity proportional to $\Omega^2$ and represents the
deviation of the equilibrium structure from spherical symmetry (see Lee \&
Baraffe (1995) for the detail of the formulation). The terms $\alpha$ and
$\beta$ respectively represent spherical expansion and rotational deformation of
the equilibrium structure. In this paper, we ignore the term $\alpha$ in
$\epsilon$ since we use evolutionary models calculated by taking account of
spherical expansion of the structure assuming uniform rotation
(e.g., Walker et al 2005).

As discussed in Lee \& Baraffe (1995), in the coordinate system $(a,\theta,\phi)$, 
the displacement vector $\pmb{\xi}$ is given by
$\pmb{\xi}=\xi^a\tilde{\pmb{e}}_a+\xi^\theta\tilde{\pmb{e}}_\theta+\xi^\phi\tilde{\pmb{e}}_\phi$, where
$\tilde{\pmb{e}}_a=(1+\epsilon+a\partial\epsilon/\partial a)\pmb{e}_r$, $\tilde{\pmb{e}}_\theta=(\partial\epsilon/\partial\theta)\pmb{e}_r+(1+\epsilon)\pmb{e}_\theta$, and $\tilde{\pmb{e}}_\phi=(1+\epsilon)\pmb{e}_\phi$, and $\pmb{e}_r$, $\pmb{e}_\theta$, and $\pmb{e}_\phi$ are orthonormal vectors in the
$r$, $\theta$, and $\phi$ directions in spherical polar coordinates.
The determinant of the metric tensor $g_{ij}$ in this coordinate system is given by
$g=\det(g_{ij})=a^4(1+\epsilon+a\partial\epsilon/\partial a)^2(1+\epsilon)^4\sin^2\theta$ and
the volume element is {given} by $dV=\sqrt{g}dad\theta d\phi$.
To the first order of $\epsilon$, we obtain $\sqrt{g}\approx a^2[1+\vartheta(\epsilon)]\sin\theta$,
where $\vartheta(\epsilon)=3\epsilon+a{\partial\epsilon/\partial a}$.

Since separation of variables is not possible to represent oscillation modes in
a rotating star, we use a finite series expansion in terms of spherical harmonic
functions for a given value of $m$, assuming axisymmetry of the equilibrium
structure. For example, the radial component of the displacement vector,
$\xi^a_{m\alpha}$, and the pressure Eulerian perturbation, $p^\prime_{m\alpha}$,
are given by
\be
{\xi^a_{m\alpha}}=\sum_{j=1}^{j_{\rm max}}S_{m\alpha, l_j}(a)Y_{l_j}^m(\theta,\phi),
\ee
and 
\be
p'_{m\alpha}=\sum_{j=1}^{j_{\rm max}}p'_{m\alpha,l_j}(a)Y_{l_j}^m(\theta,\phi),
\ee
where 
$l_j=|m|+2j-2$ for even modes, and $l_j=|m|+2j-1$
for odd modes, and $j=1,~2,~3,\cdots,~j_{\rm max}$.
Substituting perturbations represented by series expansions of finite length
into the linearized basic equations, we obtain a finite system of coupled first order linear
ordinary differential equations for the expansion coefficients, which we may call the oscillation equation.
The detail of the oscillation equation as well as the boundary conditions 
imposed at the surface and center of the star is given by Lee \& Baraffe (1995).

If we integrate {Eq.} (4) over an equipotential surface of radius $a$, we obtain
\begin{eqnarray}
\int\rho{d\ell\over dt}\sqrt{g}d\theta d\phi=
{\partial\over \partial a} F(a),
\end{eqnarray}
where 
\be
F(a)={1\over 2}a^2\sum_{m,\alpha}
{\rm Im}\left(m\pmb{S}_{m\alpha}^\dagger
\left[{\cal I}+\vartheta(\beta){\cal A}_0\right]\pmb{P}_{m\alpha}
\right),
\ee
and $\pmb{S}$ and $\pmb{P}$ are column vectors
whose $j$-th components are given by $S_{l_j}$ and $p^\prime_{l_j}$,
respectively, $\cal I$ is the unit matrix, 
and the definition of the matrix ${\cal A}_0$ is
given in Lee (1993).
Note that $\pmb{S}^\dagger=\pmb{S}^{*T}$, and $\pmb{S}^{*T}$ is the transpose of  $\pmb{S}^{*}$.
Since the quantity $F(a)$ is closely related to the work function for
oscillations in uniformly rotating stars, {acceleration
(deceleration) of the rotation velocity occurs in the damping (excitation)
regions of prograde modes}, while deceleration (acceleration) occurs in the
damping (excitation) regions of retrograde modes (e.g., Lee 2013).

It is convenient to introduce a mean $\left<f\right>$ of $f$, defined by
\be
\left<f\right>\equiv{1\over 4\pi}\int f\left[1+\vartheta(\beta)P_2(\cos\theta)\right]d\Lambda,
\ee
where $d\Lambda=\sin\theta d\theta d\phi$.
Since $\ell=r^2\sin^2\theta\Omega$ for uniformly rotating stars, we have $\left<\ell\right>
\simeq (2/3)a^2\Omega[1-(1/5)(2\beta+\vartheta(\beta))]$.
We may define the local timescale of acceleration, $\tau_s$, as
$
{\tau_s^{-1}}={\left<\ell\right>^{-1}}{\left<d\ell/ dt\right>},
$
and we have
\be
{1\over\tau_s}=
{1\over 4\pi a^2\rho\left<\ell\right>}{\partial\over \partial a}F(a),
\ee
where positive and negative $\tau_s$ respectively means local acceleration and deceleration of rotational flows around the axis.
If we assume steady mass loss, we may rewrite the left hand side of {Eq.} (8) as
\be
\int \rho a^2\left[1+\vartheta(\beta)P_2\right]v_a{\partial \ell \over\partial a}d\Lambda=
\left<\dot M{\partial\ell\over\partial a}\right>
\ee
to obtain
\be
\left<\dot M{\partial\ell\over\partial a}\right>={\partial\over\partial a}F,
\ee
where $\dot M=4\pi a^2\rho v_a$ can be regarded as a mass loss rate.
Integrating {Eq.} (13), we obtain
\be
\int_{a_0}^a\left<\dot M{\partial\ell\over\partial a}\right>da=F(a)-F(a_0).
\ee
If we  approximate $\left<\dot M\partial\ell/\partial a\right>\approx \dot M\partial \left<\ell\right>/\partial a$ and assume $\dot M$ is constant in steady state, we obtain
\be
F(a)-F(a_0)=\int_{a_0}^a\left<\dot M{\partial\ell\over\partial a}\right>da\approx J(a)-J(a_0),
\ee
where $J(a)=\dot M\left<\ell\right>$.
For the rotational acceleration to take place near the stellar surface, we need
$F(R_*)-F(a_0)>0$ for $a_0<R_*$, where $R_*$ is the radius of the star. For a
given $\dot M$, by calculating the quantity $\Delta F\equiv F(R_*)-F(a_0)$, we
may estimate the amount of angular momentum, deposited in the surface region
between $a_0$ and $R_*$, necessary to accelerate the flow from $J(a_0)$ to
$J(R_*)$.

It is convenient to normalize the quantities such as $F$, $\dot M$, and $\left<\ell\right>$ as
\be
f(a)={F(a)\over \left(GM_*^2/R_*\right)}, \quad
\overline {\left<\ell\right>}={\left<\ell\right>\over \left(R_*^2\Omega_{\rm crit}\right)},
\quad \dot m={\dot M\over \left(M_*\Omega_{\rm crit}\right)}, 
\ee
where
\be
\Omega_{\rm crit}=\sqrt{GM_*\over R_*^3},
\ee
and $G$ is the gravitational constant.
If the stellar surface at $a= R_*$ reaches the critical rotation rate $\Omega\simeq\Omega_{\rm crit}$,
we have $j\equiv \dot m\overline {\left<\ell\right>}\sim \dot m$.
If acceleration from $\Omega/\Omega_{\rm crit}\sim0.5$ to $\Omega/\Omega_{\rm crit}\simeq 1$ 
is to take place between $a_0/R_*\sim 0.95$ and the surface $a_0/R_*\simeq 1$, we have 
$j(R_*)-j(a_0)\sim 0.5\dot m$, which suggests that
we need $\Delta f\ga 0.5\dot m$ for the matter in the surface layers {to reach escape velocities above $\Omega_{\rm crit}$ 
(e.g., Meynet et al 2010) and, thus,} for a decretion disc to form around the star
as a result of angular momentum transfer by the oscillations.

\subsection{Forced oscillation calculation}

We assume that the oscillations observed in the star are stochastically excited
by the convective motions of the fluids in the convective core {and at its
boundary. This mechanism is highly complex and non-linear (e.g., Browning et al.
2004, Samadi et al. 2010, Rogers et al. 2013). Indeed, to explore the strength
of the angular momentum transport by g- and r- modes in HD\,51452, we choose to
model their stochastic excitation by turbulent convection at the core-envelope
boundary (at $r=a_b$) as resulting from periodic convective pressure
fluctuations (see also Mathis 2009) with frequencies and amplitudes that are
{constrained} thanks to the observations. This periodic
pressure {perturbation} is incorporated} as a boundary
condition and expanded as
\begin{eqnarray}
p^\prime(a_b,\theta,\phi,t)&=&p^\prime_c\Theta_{km}(\cos\theta)e^{im\phi+i\omega t}\nonumber\\
&=&p_c^\prime
\sum_lc_lY_l^m(\theta,\phi)e^{i\omega t}
\end{eqnarray}
where $\omega$ may be regarded as the frequency of the convective motions in the
core, which is related to the characteristic turn-over time $\tau_c=2\pi/\omega$
in the corotating frame {(we thus consider a perfectly resonant
excitation)}, and  $\Theta_{km}(\cos\theta)$ is the Hough function, the
eigen-solution to the Laplace's tidal equation (e.g., Lee \& Saio 1997), $c_l$
is the expansion coefficient for the function in terms of spherical harmonics, and 
we assume $k=0$ for even modes and $k=1$ for odd modes. 
In this paper, we assume that many periodic oscillations detected in the star
are excited by the convective motion of fluids in the core, and that the
frequencies detected are those  of the convective motions, which means that by
detecting many periodicities at the surface we are seeing periodic motions of
the fluids in the convective core. In this sense $p^\prime_c$ indicates the
strength (or amplitude) of the perturbations induced by the convective motions.
For a given value of $p^\prime_c$ and the frequency $\omega$, we can integrate
the oscillation equation for non-adiabatic modes to obtain the amplitude of 
$\delta L^r/L$ at the surface, which is proportional to the value of
$p_c^\prime$. Comparing $\delta L^r/L$ to the observed value, we can determine
$p_c^\prime$ and hence the amplitudes of the perturbations $\pmb{S}$ and
$\pmb{P}$, which makes it possible to estimate the amount of angular momentum
transfer taking place in the surface region of the star, using the meanflow
equation.

The oscillation equation we solve for uniformly rotating stars is the same as 
that given in Lee \& Baraffe (1995), in which global oscillation modes are
calculated as eigenmodes that satisfy the appropriate boundary conditions at the
surface and center of the stars. To calculate oscillations forced by the
convective motions of the fluids in the core, we replace the inner mechanical
boundary conditions, which are usually imposed at the stellar center, by the
condition given by {Eq.} (18) imposed at the core-envelope boundary. The
thermal boundary condition at the boundary is that the oscillations are
adiabatic in the deep interior. The surface boundary conditions are the same as
those given in Lee \& Barrafe (1995). The replacement of the inner mechanical
boundary conditions makes the system of linear differential equations
inhomogeneous, which makes it possible to integrate the system of differential
equations for arbitrary forcing frequencies $\omega$.  We solve the oscillation
equation between the core-envelope boundary and the stellar surface, and we
employ the Cowling approximation for simplicity.

\section{Numerical results}

\begin{figure}
\resizebox{0.9\columnwidth}{!}{
\includegraphics{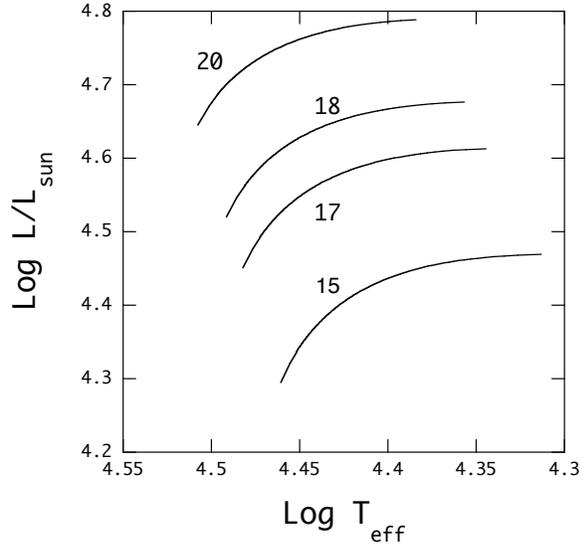}}
\caption{Evolutionary tracks from ZAMS of rotating main sequence stars of  15,
17, 18, and 20 $M_\odot$ models for $X=0.7$ and $Z=0.02$. Uniform rotation of
the  frequency 1.22 cycle per day is assumed during the evolution.}
\end{figure}

We calculate forced oscillations for the observed oscillation frequencies and
luminosity amplitudes tabulated in Neiner et al. ({2012a}) for main sequence stars.
We compute the main sequence models using a standard stellar evolution code with
a OPAL opacity table for $X=0.7$ and $Z=0.02$, and we incorporate the spherical
expansion due to uniform rotation by replacing the {usual} hydrostatic
balance equation $dp/dr=-\rho GM_r/r^2$ by $dp/dr=-\rho
GM_r/r^2+(2/3)r\Omega^2$, {where the average on latitude of the centrifugal
acceleration is taken into account}. We use the observed rotation rate 1.22
cycle per day for the model calculation, and since we fix {this} rate as a
constant, the ratio $\Omega/\Omega_{\rm crit}$ increases with
{the} evolution of the {star} from
{the} ZAMS due to the radial expansion and exceeds unity during
the main sequence stage{. We stop} the evolution calculation for
$\Omega/\Omega_{\rm crit}\ga 1$. Figure 1 shows evolutionary tracks from
{the} ZAMS of uniformly rotating stars of different masses. As
a fiducial model for the analysis in this paper, we use a rotating main sequence
star model of $18M_\odot$, whose physical parameters are  
$\log (L/L_\odot)=4.6049$, $\log T_{\rm eff}=4.4691$,
$R_*/R_\odot=7.1624$, $\log g=3.9829$, and $\bar\Omega\equiv \Omega/\Omega_{\rm
crit}=0.6425$ for the rotation frequency 1.22 cycle per day, where $\Omega_{\rm
crit}=1.381\times10^{-4}~\rm s^{-1}$. 
These physical parameters for the model are consistent with those
cited in Neiner et al. (2012a) for the Be star HD$~$51452.
For this {model}, we have $\dot
m=1.26\times 10^{-15}$ for $\dot M=10^{-10}M_\odot/\rm yr$.

\begin{figure*}
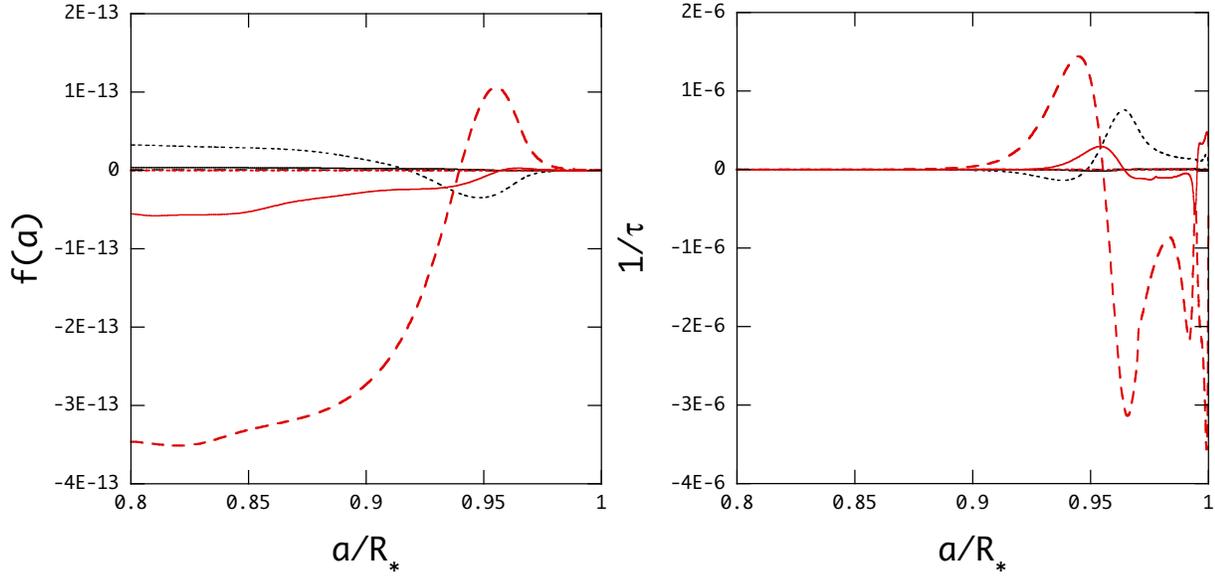

\resizebox{0.45\textwidth}{!}{
\includegraphics{f2a.epsi}}
\resizebox{0.45\textwidth}{!}{
\includegraphics{f2b.epsi}}
\caption{$f(a)$ and $1/\tau$ versus $a/R_*$ for forced oscillations of even parity for $m=1$, where the solid, dashed, and dotted curves are for the forced frequencies
$|\bar\omega|=0.5$, 1, and 1.5, respectively, and the black and red curves are
for retrograde and prograde waves, respectively. Here,
$\bar\omega=\omega/\Omega_{\rm crit}$, and the time-scale $\tau$ is normalized by $\Omega_{\rm crit}^{-1}$.}
\end{figure*}

\begin{figure*}
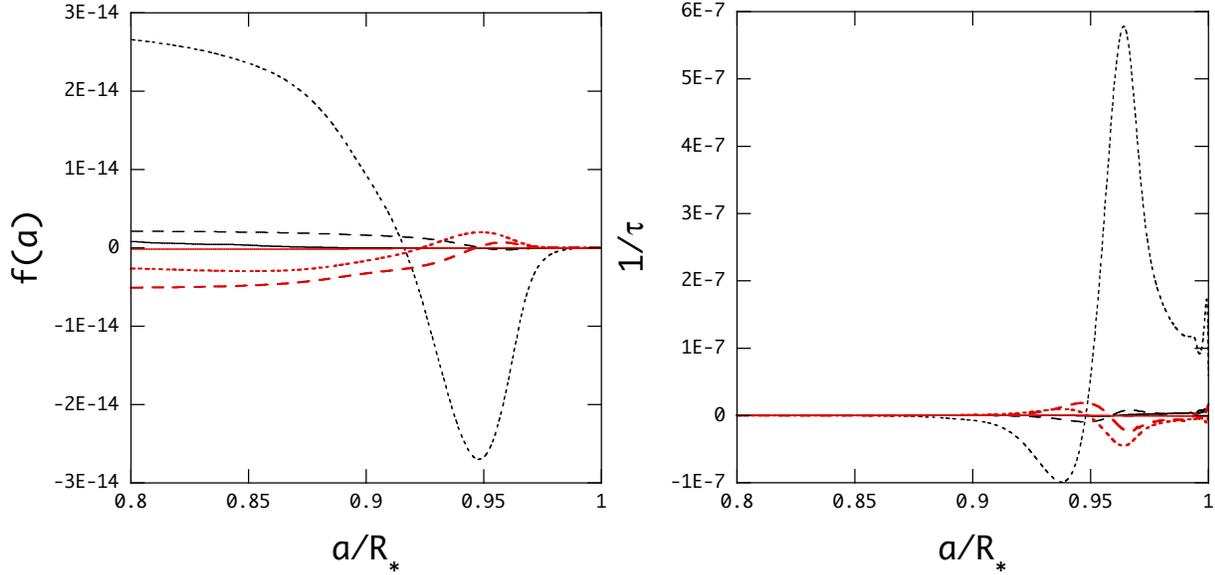

\resizebox{0.45\textwidth}{!}{
\includegraphics{f3a.epsi}}
\resizebox{0.45\textwidth}{!}{
\includegraphics{f3b.epsi}}
\caption{Same as Figure 2 but for forced oscillations of odd parity. }
\end{figure*}

The observation suggests that the star is rapidly rotating, as rapidly as
$\Omega/\Omega_{\rm crit}\ga 0.5$, but most of the detected frequencies in the
observer's frame are small compared to the critical frequency $\Omega_{\rm
crit}$ of the star. This indicates that most of the frequencies may be
attributable to low-frequency {gravito-inertial waves} in the corotating
frame, and that the effects of rotation on the oscillations become crucial,
particularly for the oscillations whose frequencies in the corotating frame are
comparable to or {lower} than the rotation frequency $\Omega$.

As is well {known}, the opacity bump due to iron group elements
in the temperature regions of $T\sim 2\times 10^5$K in the stellar interior can excite
high radial order $g$ modes in slowly pulsating B (SPB) stars and low radial
order $p$ modes in $\beta$ Cephei stars. Although the opacity bump at $T\sim
2\times 10^5$K cannot be effective enough to excite low-frequency modes in main
sequence stars of $M\sim 18M_\odot$ such as HD\,51452, it is expected that {this
high-$\kappa$} region located close to the stellar surface may contribute to
angular momentum transfer in the surface layers of the star {by} global
oscillations excited by convective motions {in} the core {because of
their radiative damping (e.g., Zahn, Talon, Matias 1997) as well as 
$\kappa$-driving due to the opacity bump (e.g., Lee 2013).} 
Since angular momentum deposition to (extraction from) the meanflow {depends on
the sign of $m$},
it is important to know whether the observationally identified oscillations are
prograde or retrograde in the corotating frame of the star when we sum up all
contributions from the modes considered.

\begin{figure*}
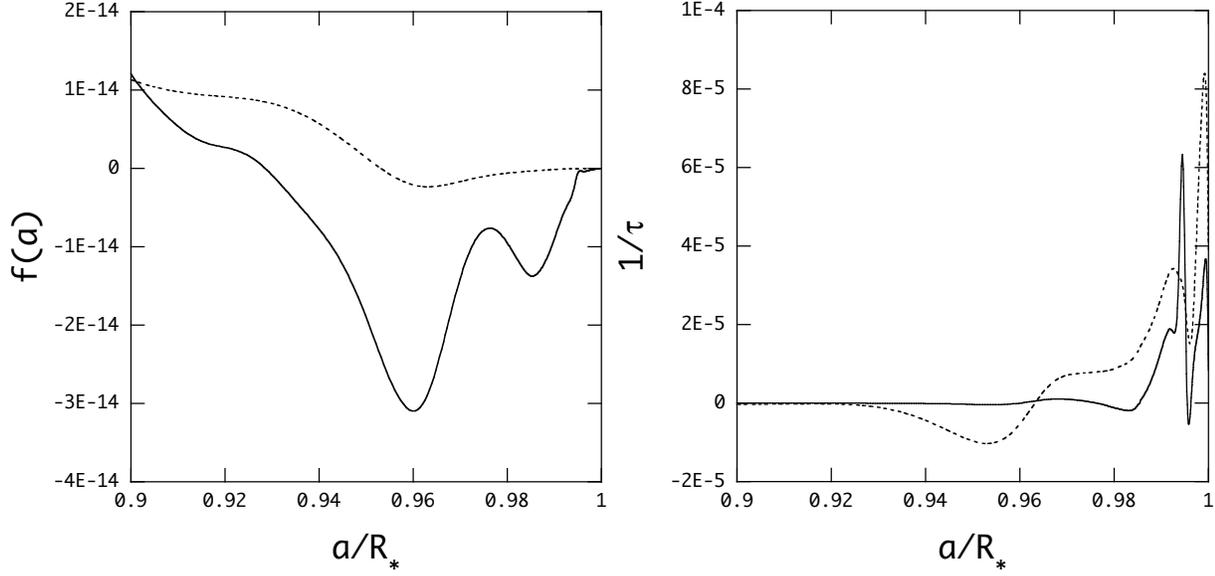

\resizebox{0.45\textwidth}{!}{
\includegraphics{f4a.epsi}}
\resizebox{0.45\textwidth}{!}{
\includegraphics{f4b.epsi}}
\caption{$f(a)$ and $1/\tau$ versus $a/R_*$ for $m=1$ waves for the forcing
frequency $\bar\omega=2m(\bar\Omega-\delta)/[l^\prime_1(l^\prime_1+1)]$ with
$\delta=10^{-3}$, where $l^\prime_j=l_j+1$ for even parity mode (solid line) and
$l^\prime_j=l_j-1$ for odd parity mode (dotted line), and $\tau$ is normalized
by $\Omega_{\rm crit}^{-1}$. We plot $(1/\tau)\times 100$ for the odd parity
wave. Note that $2m\bar\Omega/[l^\prime_1(l^\prime_1+1)]$ is an asymptotic
frequency of $r$-modes.}
\end{figure*}

\begin{figure*}
\resizebox{0.45\textwidth}{!}{
\includegraphics{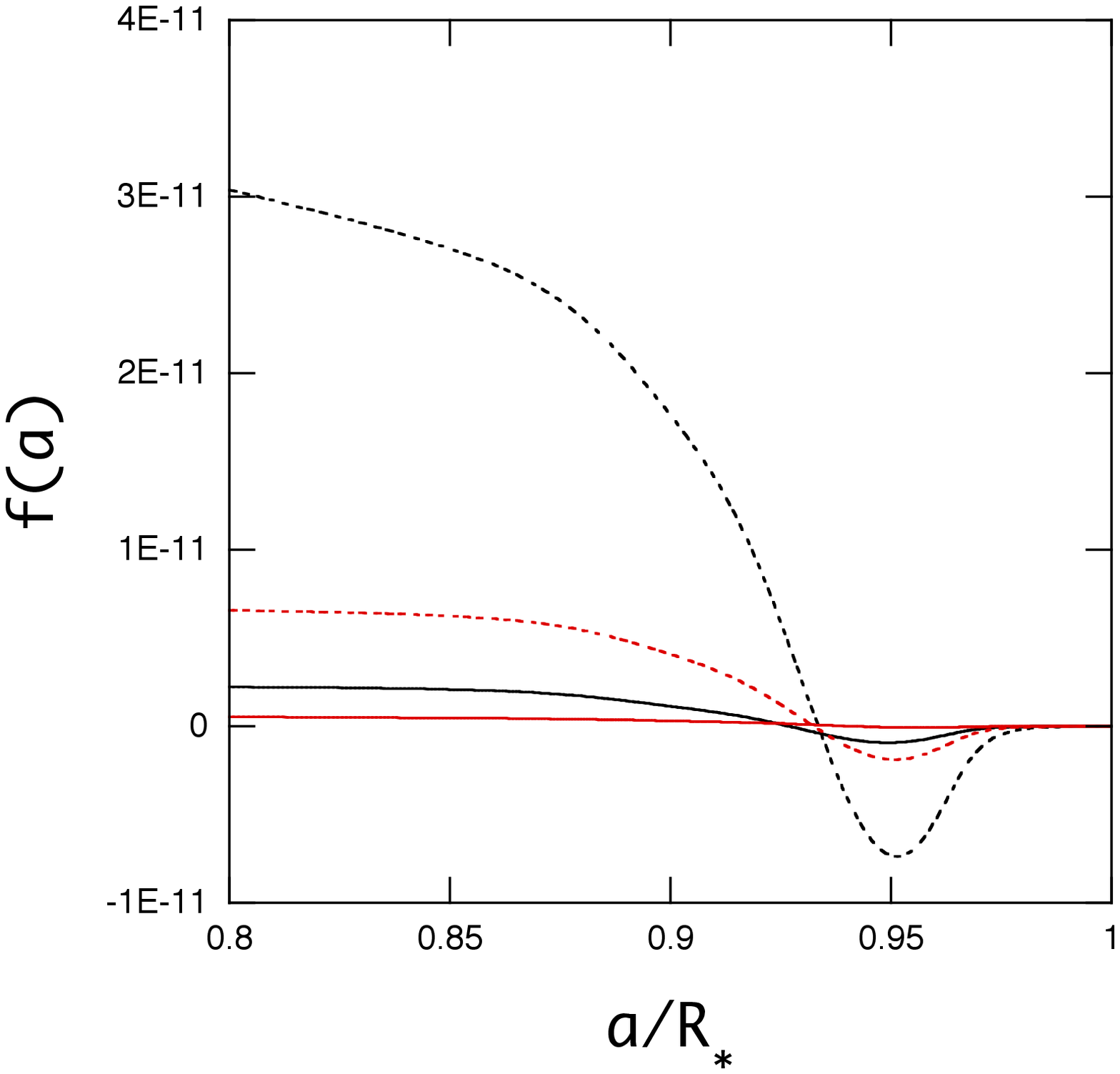}}
\resizebox{0.45\textwidth}{!}{
\includegraphics{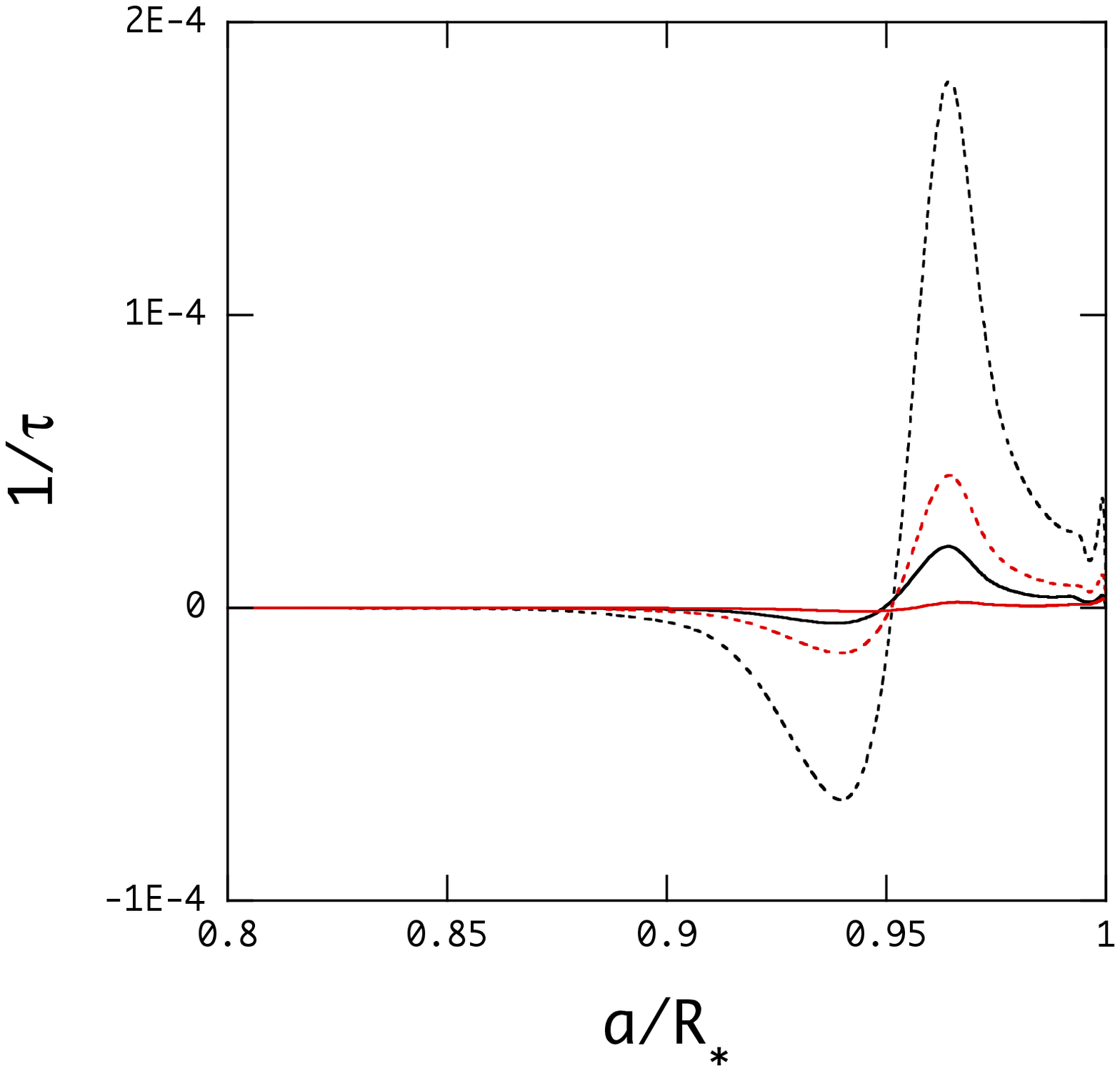}}
\caption{$f(a)$ and $1/\tau$ versus $a/R_*$ for $m=1$ (solid lines) and 2
(dotted lines) calculated by summing up contributions from all the forcing
frequencies identified by Neiner et al. ({2012a}) for {the} hot Be
star HD\,51452, where the black lines and red lines are for even parity waves of
$l=|m|$ and odd parity waves of $l=|m|+1$. }
\end{figure*}

\begin{figure*}
\resizebox{0.45\textwidth}{!}{
\includegraphics{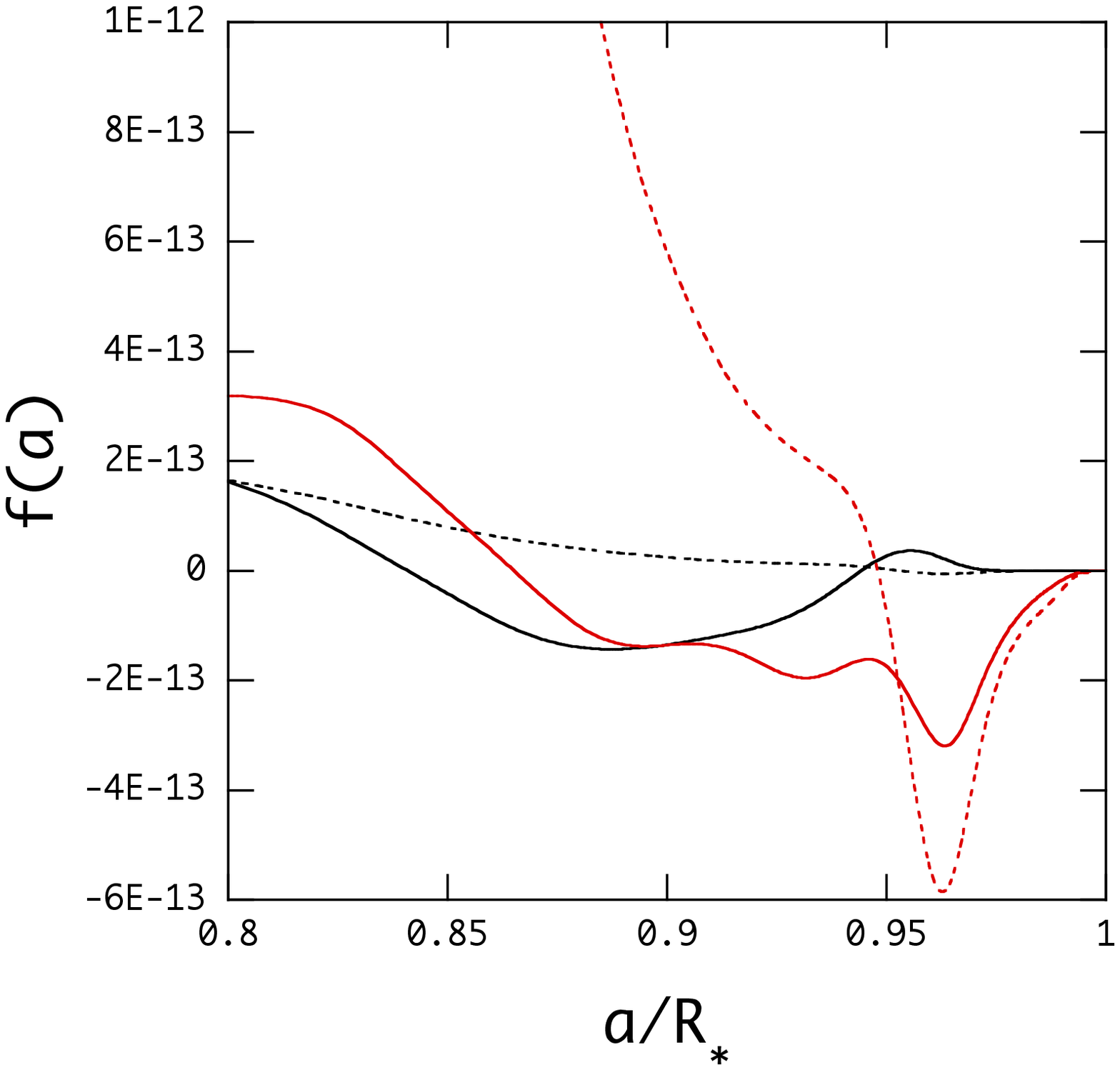}}
\resizebox{0.45\textwidth}{!}{
\includegraphics{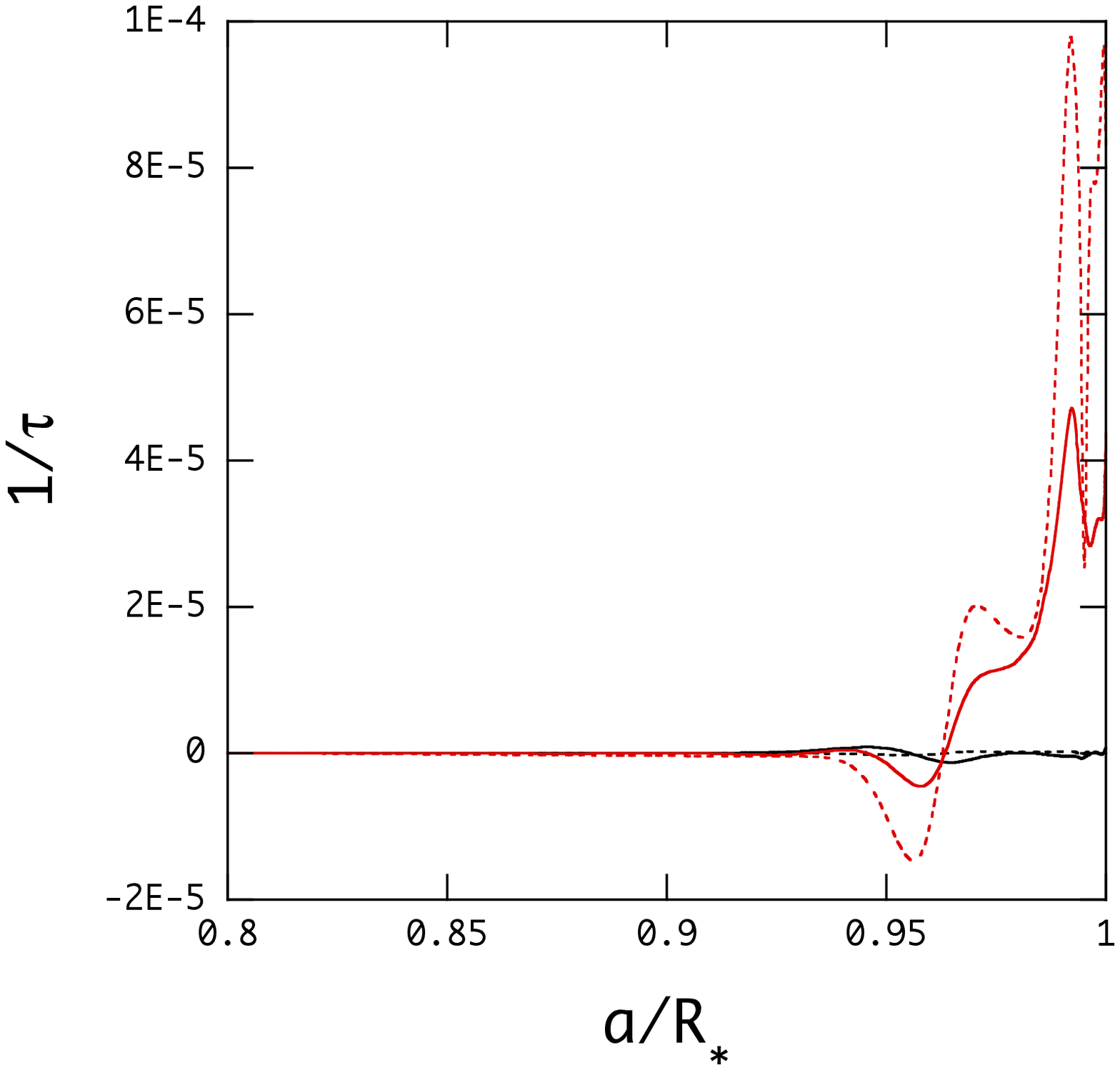}}
\caption{Same as Figure 5 but for $m=-1$ and $-2$.
}
\end{figure*}

Assuming forced oscillations, we can calculate the quantities $f(a)$ and
$1/\tau$ for arbitrary frequencies. For the case of $m=1$, we plot $f(a)$ and
$1/\tau$ in Figure 2 (even parity waves) and Figure 3 (odd parity waves) for the
forced frequencies $|\bar\omega|=0.5$, 1, and 1.5 for prograde (red curves) and
retrograde (black curves) waves, where $\bar\omega\equiv \omega/\Omega_{\rm
crit}$. Note that the amplitudes of the forced oscillations are determined so
that the calculated $(\delta L^r)_{l_1}/L$ at the surface is equal to the
largest observed amplitude mode having the frequency 0.6293 cycle per day. The
rapid changes in the function $f(a)$ occurs in the surface regions, and hence
the quantity $1/\tau$, which is quite small in the deep interior, has finite
values in the surface layers. The functions $f(a)$ and $1/\tau$ for prograde and
retrograde waves are roughly symmetric about the $a-$axis. We note that
{the} strong {high-$\kappa$ region,} due to the opacity
bump produced by iron group elements, is located at $a/R_*\simeq 0.95$, and that
there appears a strong damping region below {this} region. As
indicated by the right panels of Figs. 2 and 3 for $1/\tau$, a strong acceleration (deceleration) by
retrograde (prograde) oscillations takes place in the driving
region due to the opacity bump, while a strong deceleration (acceleration) by retrograde (prograde)
oscillations takes place in the damping region. It is also interesting to note
that the amount of the difference $\Delta f$ between $a/R_*=0.95$ and 1 can be
of order of $10^{-14}$ for the frequency $\bar\omega\simeq 1.5$ for both
prograde and retrograde oscillations and {that} it is larger than $\dot
m\sim 10^{-15}$ for $\dot M=10^{-10}M_\odot~{\rm yr}^{-1}$. This may suggest
that  even a single mode can transfer an amount of angular momentum large 
enough for a decretion disc to form around the star. Of course, if many modes
are excited at the same time by the external force, 
we have to sum up all contributions from the {excited} modes to both
acceleration and deceleration in order to obtain the net amount of angular
momentum transferred to the surface layers of the star.

Figure 4 also plots the functions $f(a)$ and $1/\tau$ of $m=1$ waves for the
forcing frequencies
$\bar\omega=2m(\bar\Omega-\delta)/[l^\prime_1(l^\prime_1+1)]$ with
$\delta=10^{-3}$,  which are close to an asymptotic frequency of $r$-modes,
where $l^\prime_j=l_j+1$ for even modes and $l^\prime_j=l_j-1$ for odd modes.
$R$-modes are retrograde modes and angular momentum deposition occurs in the
layers much closer to the surface than $g$-modes. The amount of angular momentum
$\Delta f$ transferred by the $r$-modes to the surface layers can be comparable
to $\dot m\sim 10^{-15}$.

To estimate the possible amount of angular momentum transferred by oscillations
excited by convective motions of the core fluid, we have to sum up contributions
from all the modes. As shown in the previous paragraphs, however, for a correct
estimation we need to know, besides the frequencies for forcing, the value of
$m$ for each of the modes. This sort of information, however,
{could not be determined from the observations} in Neiner et
al. ({2012a}). In this paper, therefore, to calculate forced oscillations having
the detected frequencies, we assume, as a working hypothesis, that the observed
oscillations are all prograde waves or all retrograde waves in
{the} observer's frame,  having a single value of the wave
number $m$.  Since $\omega=\sigma+m\Omega$, if the observed oscillations are all
retrograde waves in the observer's frame such that $m>0$ and $\sigma>0$,  we
have $\omega>0$ for all the oscillations, that is, the oscillations are all
retrograde waves in the corotating frame. On the other hand, if the oscillations
are all prograde waves in the observer's frame such that $m<0$ and $\sigma>0$,
the oscillation frequency $\omega$ in the corotating frame can be both positive
(prograde waves) and negative (retrograde waves). Note that almost all the
oscillations could be retrograde for large $|m|$ when $\Omega\sim \Omega_{\rm
crit}$ and $\sigma\la\Omega_{\rm crit}$.

Since the observed oscillations should have amplitudes large enough to be
detected at the stellar surface by the {CoRoT satellite}, it is
reasonable to rule out the oscillations that have large amplitudes only in the
deep interior, and such confinement usually takes place for low-frequency
oscillations in the corotating frame for given values of $m$ and $l$.
{To} exclude such low-frequency waves confined in the deep
interior, defining the kinetic energy of the oscillation as
$E_k(a)=\omega^2\int_{a_b}^{a}\rho\xi^*\cdot\xi dV$, we pick up only the
oscillations for which $E_k(0.5R_*)/E_k(R_*) < 0.9$.
{The} oscillations ruled out for a given set of $m$
and $l$ are not necessarily ruled out for other combinations of $m$ and $l$.

{We chose to use $|m|$=1 and 2, since these $|m|$ values are those typically identified in observations of Be stars (e.g., Rivinius et al. 2003).}

In Figure 5, we plot the functions $f(a)$ and $1/\tau$ for the observed
frequencies for $m=1$ and 2 for both even and odd parities.  Since the detected
oscillations are all assumed to be retrograde in the corotating frame,  the
contributions coming from the waves' driving 
zone due to the opacity bump
close to the stellar surface all work for increasing the rotation velocity. 
We also note that the amount of angular momentum deposition there, $\Delta f$, is
much larger than $\dot m\sim10^{-15}$ for $\dot M=10^{-10}M_\odot/{\rm yr}$,
which suggests that the amount of angular momentum deposition is large enough
for a decretion disc to form around the star. We have to keep in mind that the
estimation of the angular momentum deposition here could be simply an over
estimation.

Figure 6 is for $m=-1$ and $-2$ and in this case the oscillations in the
corotating frame can be both prograde and retrograde. Since
{the high opacity}  zone works for accelerating (decelerating)
{the} rotation velocity for retrograde (prograde) waves in the
corotating frame, the net effects of angular momentum deposition at a given
radius $a$ are determined by the sum of {the} contributions
from all the retrograde and prograde waves, and this summation  leads to a
rather complicated behavior of the quantities $f(a)$ and $1/\tau$ as a function
of $a/R_*$ in the surface regions. The amount of angular momentum deposition
$\Delta f$ in the interval of $0.95\la a/R_*\la 1$ is of {the} order of $10^{-13}$ for
odd parity oscillations and is larger than {the} value of $\dot m\sim 10^{-15}$
necessary for a disc formation with 
$\dot M\sim 10^{-10}M_\odot/{\rm yr}$.

The minimum timescale $\tau$ of acceleration in the surface regions is attained
at $a/R_*\simeq 0.96$ for positive $m$ ({Fig.} 5) and at $\sim 0.99$ for
negative $m$ ({Fig.} 6). The minimum timescale $\tau$ is of order of $10^{4}$
for both cases and since $\Omega_{\rm crit}\sim 10^{-4}~{\rm s^{-1}}$, the acceleration
timescale is of order of a few years. {This timescale is typical of the recurrence timescale of outbursts in Be stars.}

{
\section{Conclusions}
}

Using a theory of wave-meanflow interaction, we have estimated the possible
amount of angular momentum transferred by {gravito-inertial waves} having
the set of frequencies observationally detected for {the} Be
star {HD\,51452}. Since the Be star is rapidly rotating and the
detected frequencies are low in the observer's frame in the sense that
$\sigma\la\Omega_{\rm crit}$, the frequencies $\omega$ in the corotating frame
of the star {are} low, that is, the detected frequencies are in
the frequency domain of $g$-modes and $r$-modes. Since the opacity
{$\kappa$} mechanism does not work for excitation of low
frequency modes in massive stars of $M_*\ga 10M_\odot$ {for a
solar metallicity (Pamyatnykh 1999, Miglio et al. 2007)}, we need to look for an
alternative mechanism for the excitation of the observationally detected low
frequency oscillations. In this paper, we assume that the oscillations are
excited by periodic convective motions of the core fluid in the massive star
{as suggested by Neiner et al. (2012a)}. We treat therefore the
detected oscillation frequencies as forcing frequencies for waves propagating in
the radiative envelope. We calculate the forced oscillations having the observed
frequencies by imposing pressure perturbations produced by the convective fluid
motions {at the core-envelope interface as the boundary condition}. This
procedure makes inhomogeneous the system of linear differential equations for
oscillations, which can be integrated for arbitrary frequencies.

In the theory of wave-meanflow interaction, {the high-$\kappa$} regions
works for accelerating
(decelerating) the rotation (mean)flows for retrograde (prograde) waves observed
in the corotating frame of the star. Since we have no information
{on} the azimuthal wave number $m$ for the detected
oscillations, we calculated the oscillations of the frequencies just assuming
that they are all retrograde modes or prograde modes in the observer's frame.
{In} the former case, all the modes are retrograde in the
corotating frame and hence the angular momentum deposition can take place
efficiently in the surface layers of the star. {In} the latter
case, however, the modes can be separated into prograde and retrograde modes,
which means that even in {the high-$\kappa$} regions
the net amount of
angular momentum deposition depends on the net sum of accelerating and
decelerating contributions of retrograde and prograde waves.

{Imposing the observed} frequencies, we have shown that the
amount of angular momentum transferred to the surface regions is large enough
for decretion disc formation with the mass loss rate $\sim10^{-10}M_\odot/\rm
yr$, although a definite amount of  angular momentum transferred can be
estimated only after the details {of mode identification are
determined. This could possibly be done with a ground-based high-resolution
spectroscopic campaign.}

Since we assume forced oscillations propagating in the radiative envelope of the
star, there exist the possibility that some of the forcing frequencies, {
that are determined by properties of motions in the convective core}, are in
resonance with {eigen}frequencies of the {free} oscillation modes such
as $g$-modes or $r$-modes in the envelope. Both the forcing frequencies and the
{eigen}frequencies can change with {the} {star's}
evolution. If {such} a frequency resonance {happens}
between an {envelope} {free} mode and a forcing frequency,
{its} {amplitude} can be increased so that the effects of
angular momentum deposition or extraction in the surface regions is enhanced.

{In this paper, we have demonstrated, using the case of HD\,51452, the
ability of gravito-inertial waves to transport angular momentum efficiently
{in} rapidly rotating massive stars} {and to 
deposit this momentum just below the surface. This mechanism increases the
velocity of the surface layers. These layers can then reach the escape velocity
at which material gets ejected from the star. Therefore, this mechanism could be
at the origin of matter ejections in Be stars and the reason of the presence of
a decretion disc around Be stars.}

{In a near future, we will improve the physical modelling of this
wave-meanflow. First, in this work, we have assumed a uniform rotation to
compute the oscillations. However, because angular momentum deposition occurs in
a rather short time-scale in the surface layers, it may be important to take
{differential rotation into account} when computing modes and
the transport of angular momentum they {induce} (e.g., Lee \&
Saio 1993; Mathis 2009). In this way, we will get a complete picture of
wave-meanflow interactions. Next, the modelling of the amplitude of waves, which
are stochastically excited by the convective core in massive stars, must be
improved using theoretical models (e.g., Belkacem et al. 2009; Lecoanet \&
Quataert 2013; Mathis, Neiner \& Tran Minh 2014) and {more and
more realistic} direct numerical simulations (e.g. Browning et al. 2004, Rogers
et al. 2013). Besides, differential rotation and related meridional flows and
turbulence play an important role in the evolution of massive stars (e.g.,
Meynet \& Maeder 2000). In this context, it would be important to take these
mechanisms {into account as they} also transport angular
momentum simultaneously with gravito-inertial waves. This has
{already} been done for the case of solar-type stars (Talon \&
Charbonnel 2005, Charbonnel et al. 2013, Mathis et al. 2013), in which waves
strongly modify the star's rotational evolution, {but} the case
of massive stars {still has} to be studied. Such study
{should} be undertaken first in spherical models
{in which} the centrifugal acceleration is treated as a
perturbation to unravel the role of each physical processes. Then, complete 2D
models must be developped for rapidly rotating stars (e.g., Ballot et al. 2010;
Espinosa Lara \& Rieutord 2013).}

\bigskip

\section*{Acknowledgments}
This work was supported in part by the Programme National de Physique Stellaire
(CNRS/INSU), the ANR SIROCO, the CNES/CoRoT grant at the LESIA (Observatoire de
Paris) and Service d'Astrophysique (CEA Saclay), and the CNRS Physique
th\'eorique et ses interfaces program.

\end{document}